\documentclass{aastex6}

\def\div{{\nabla\cdot\,}}	% divergence
\def\half{\frac{1}{2}}
\def\3half{\frac{3}{2}}

\begin{document}

\title{
Numerical Simulation of Star Formation by the Bow Shock of the Centaurus A Jet}

\author{
Carl~L. Gardner\altaffilmark{1},
Jeremiah~R. Jones\altaffilmark{1}}
\altaffiltext{1}{School of Mathematical \& Statistical Sciences,
Arizona State University, Tempe AZ 85287,
carl.gardner@asu.edu, jrjones8@asu.edu}

\author{Evan Scannapieco\altaffilmark{2}}
\and
\author{Rogier~A. Windhorst\altaffilmark{2}}
\altaffiltext{2}{School of Earth \& Space Exploration,
Arizona State University, Tempe AZ 85287,
evan.scannapieco@asu.edu, rogier.windhorst@asu.edu}

\begin{abstract}

Recent {\em Hubble Space Telescope (HST)}\/ observations of the
extragalactic radio source Centaurus~A (Cen~A) display a young stellar
population around the southwest tip of the inner filament 8.5 kpc from
the Cen~A galactic center, with ages in the range of 1--3 Myr.
Crockett et al.\ (2012) argue that the transverse bow shock of the
Cen~A jet triggered this star formation as it impacted dense molecular
cores of clouds in the filament.  To test this hypothesis, we perform
three-dimensional numerical simulations of induced star formation by
the jet bow shock in the inner filament of Cen~A, using a positivity
preserving WENO method to solve the equations of gas dynamics with
radiative cooling.  We find that star clusters form inside a
bow-shocked molecular cloud when the maximum initial density of the
cloud is $\ge 40$ H$_2$ molecules/cm$^3$.  In a typical molecular
cloud of mass $10^6 \, M_\odot$ and diameter 200 pc, approximately 20
star clusters of mass $10^4 \, M_\odot$ are formed, matching the {\em
  HST}\/ images.

\end{abstract}


\section{Introduction}

The nearest powerful extragalactic radio source is Centaurus~A
(Cen~A), which is emitted by the active galactic nucleus (AGN) of the
elliptical galaxy NGC 5128.  Unlike most elliptical galaxies, NGC 5128
hosts a prominent dust lane that is thought to have been created by a
recent merger with a gas-rich disk galaxy (Israel 1998), which also
provided the fuel for the AGN outburst.  At a distance of 3.7 Mpc from
the Earth, Cen~A provides a unique opportunity to study the evolution
of AGN driven radio jets.

Centaurus A is at the center of a relatively low mass $\approx 2
\times 10^{12} M_\odot$ galaxy subgroup (Karachentsev et al.\ 2002),
one of two subgroups within the Cen~A/M83 galaxy group.  The subgroup
spans a region $\approx 400$ kpc in radius around NGC 5128, which at
this close distance covers many degrees in the southern sky.  Cen~A is
similarly impressive at radio wavelengths, showing multiple
synchrotron emitting lobes, ranging in scale from a pair of inner
lobes at a distance of 5 kpc from the AGN, to a single Northern Middle
Lobe at a distance of 25--30 kpc, and to a pair of outer lobes at a
distance of 100 kpc from the nucleus (Morganti et al.\ 1999, Saxton et
al.\ 2001, Feain et al.\ 2009).  Furthermore, the bent morphologies of
the jets, as well as their association with X-ray emitting knots and
regions of bright optical and UV emission, all suggest a complex
history of interactions between the jets and the surrounding gaseous
environment (Rejkuba et al.\ 2001, 2002, 2004, Fassett \& Graham 2000,
Graham \& Fassett 2002, Kraft et al.\ 2009, Gopal-Krishna \& Wiita
2010).

Of particular interest is the connection between the radio features
and star formation.  Such AGN feedback is widely invoked in
theoretical models both to explain the overall history of cosmic star
formation, as well as to explain the properties of individual AGN host
galaxies and their surroundings.  On cosmic scales, it is thought that
the primary impact of AGN is to decrease star formation by removing
gas from galaxies and heating the gas that surrounds them.  Such
``negative feedback'' is required to explain three key cosmological
trends: (i) that the overall star formation rate has dropped since $z
\approx 2$, 4 Gyr after the Big Bang (Hopkins \& Beacom 2006, Karim et
al.\ 2011); (ii) that the rate of star formation peaked first in the
most massive galaxies and later in smaller galaxies (e.g., Cowie et
al.\ 1996, Brinchmann \& Ellis 2000, Noeske et al.\ 2007, Cowie \&
Barger 2008); and (iii) that galaxy clusters that show rapid gas
cooling in their centers do not show evidence of strong star formation
(e.g., Peterson et al.\ 2001, Rafferty et al.\ 2006, Cavagnolo et
al.\ 2009).  Negative AGN feedback in the form of radiatively driven
winds and radio jets has been shown to address these questions by (i)
providing enough energy to remove and heat the gas that would
otherwise form stars below $z = 2$ (Granato 2004, Scannapieco \& Oh
2004, Croton et al.\ 2006); (ii) preferentially reducing star
formation first in large galaxies at early times, and then in smaller
galaxies at late times (e.g., Scannapieco \& Oh 2004, Scannapieco et
al.\ 2005, Hopkins et al.\ 2006, Somerville et al.\ 2008, Johansson et
al.\ 2012, Wurster \& Thacker 2013, Costa et al.\ 2014, Richardson et
al.\ 2016); and (iii) providing the energy to reheat the cooling gas
in cold-core galaxy clusters (e.g., Dunn \& Fabian 2006, McNamara \&
Nulsen 2007, 2012, Fabian 2012, Gaspari et al.\ 2012a, 2012b, Li et
al.\ 2015).

In more limited cases, ``positive feedback'' from active galactic
nuclei may induce star formation by creating expanding rings of star
formation around the centers of galaxies, triggered by jet bow shocks
(see Fragile et al.\ 2004, Gaibler et al.\ 2012, Silk 2013, Zinn et
al.\ 2013, Dugan et al.\ 2014, Gaibler 2014, Wagner et al.\ 2015,
Dugan et al.\ 2016).  At $z = 3.9$, the galaxy 4C 41.17 has been
observed to host a powerful radio jet with strong evidence for induced
star formation along the radio axis (Dey et al.\ 1997, Bicknell et
al.\ 2000).  At $z =0.31$, multi-wavelength observations of the radio
source PKS2250-41 suggest that the radio jet in this source has
triggered recent star formation within a faint companion. In the
galaxy cluster Abell 194, which is 78.7 Mpc away from the Earth, the
peculiar galaxy ``Minkowski's Object'' (Minkowski 1958) is likely to
be experiencing a burst of star formation triggered by the radio jet
from the nearby active galaxy NGC 541 (van Breugel et al.\ 1985, Croft
et al.\ 2006).

\begin{figure}[htbp]
\begin{center}
\includegraphics[scale=0.4]{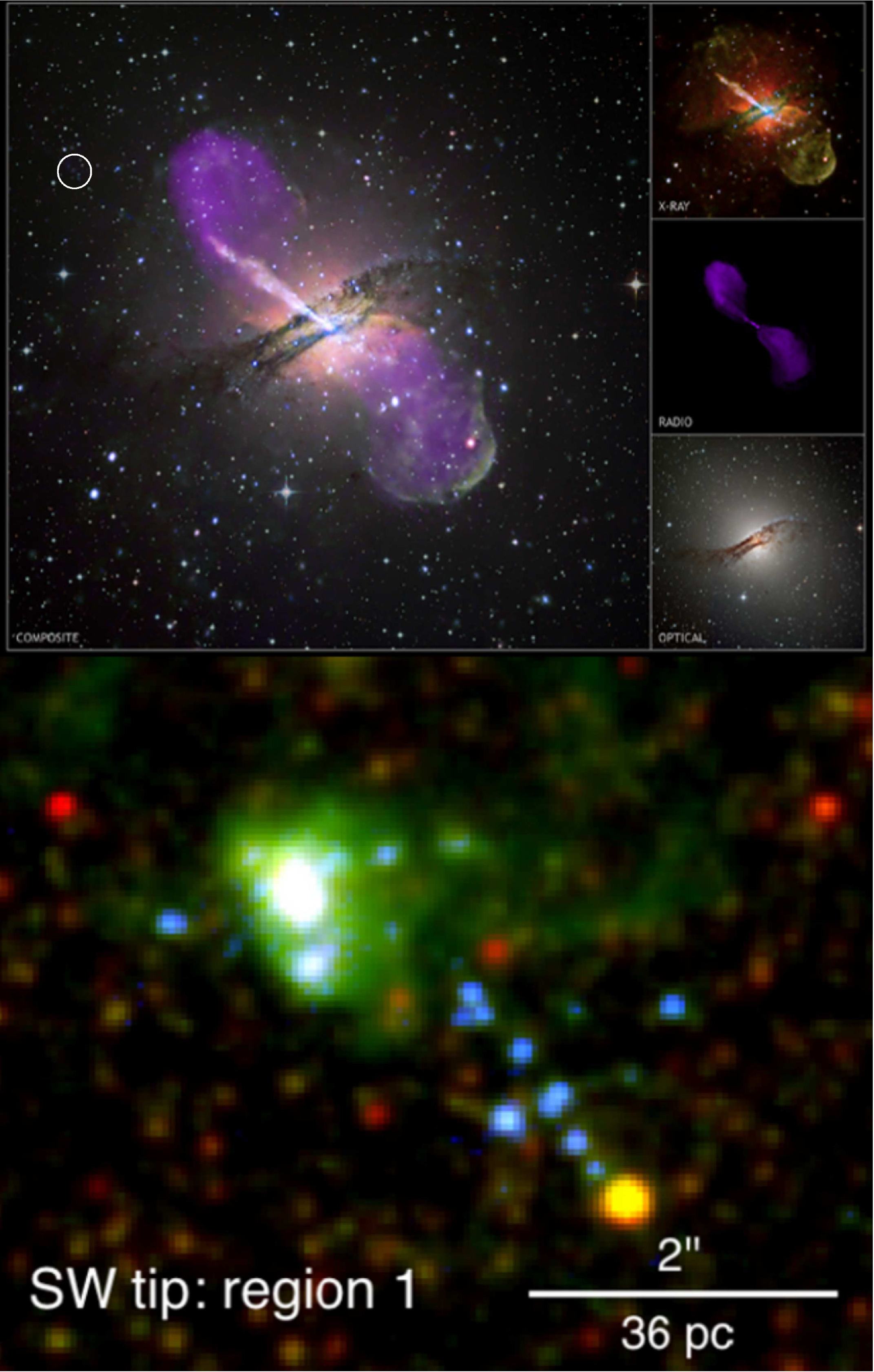}
\caption{Top panel: Composite of Chandra X-ray image, VLA radio image,
  and ESO optical image of the Cen~A jet and galaxy NGC 5128: X-ray
  image NASA/CXC/CfA/R.~Kraft et al.; radio image
  NSF/VLA/Univ.~Hertfordshire/M.~Hardcastle; optical image
  ESO/WFI/M.~Rejkuba et al\@.  A 1 kpc diameter area surrounding
  cloud~A (which is about 200 pc in diameter and contains Region~1) in
  the inner filament is marked by a white circle.  Bottom panel: A
  zoomed-in {\em HST}\/ view of Region~1 (Crockett et al.\ 2012),
  where numerous young star clusters can be seen as bright blue dots.
  Region~1 is estimated to contain about $10^5 \, M_\odot$, while the
  individual clusters are estimated to contain about $10^3 \, M_\odot$
  each.}
\label{fig:CenAStars}
\end{center}
\end{figure}

In the case of Cen~A, two optical filaments with recently formed star
clusters are observed near the northern radio jet.  Recent {\em Hubble
  Space Telescope (HST)}\/ observations by Crockett et al.\ (2012) of
the inner filament of Cen~A display a young stellar population around
the southwest tip of the filament (reproduced in
Figure~\ref{fig:CenAStars} here), with ages in the range of 1--3 Myr.
Crockett et al.\ (2012) argue that the transverse bow shock of the
Cen~A jet triggered this star formation as it impacted the dense
molecular cores of three clouds in the filament.  If this view is
correct, then the inner filament of Cen~A would provide one of the
nearest and best measured examples of the positive feedback of AGN
activity on the subsequent formation of stars.
 
As indicated in Figure~\ref{fig:CenAStars}, the jet-induced star
formation in the inner filament occurs at some distance from the
visible, radio, and X-ray portions of the northern jet, more or less
along the (invisible) extension of the X-ray jet (see Figure~1 in
Crockett et al.\ 2012).  Note that this latter figure implies that the
physical conditions required in the simulations to trigger star
formation still hold this far away from the jet, because the inner
filament is near the edge of the radio jet.  In other words, the jet
bow shock is still expanding laterally beyond the region where the jet
is visible in radio and X-ray emission.  The AGN jet may also be
precessing around the central engine on a timescale of approximately
$10^7$ years, but the effects of the precession are beyond the scope
of the current paper.

Sutherland et al.\ (1993) suggested that radio jet-induced shocks can
produce the observed optical line emission from the inner filament.
However, Crockett et al.\ (2012) maintain that such shocks are
produced by the bow shock propagating through the diffuse interstellar
medium, rather than by the radio jet directly.  As the bow shock
collides with molecular clouds precipitating star formation in the
dense molecular cores, ablation and heating of the diffuse gas result
in the observed optical line and X-ray emission.  Wagner et
al.\ (2015) found that star formation occurs in dense clouds with
diameters greater than about 25 pc, but that bow shock ablation
quenches star formation in smaller clouds.

This investigation focuses on jet-induced localized star formation in
Cen~A in larger molecular clouds with diameters between 130--325 pc,
far away (8.5 kpc) from the AGN.  To test the hypothesis, we perform
three-dimensional (3D) numerical simulations of induced star formation
by the jet bow shock in the inner filament of Cen~A, using a
positivity preserving WENO method to solve the equations of gas
dynamics with radiative cooling.  The number of star clusters formed,
as well as their sizes and masses, are in good agreement with the {\em
  HST}\/ observations.

The structure of this investigation is as follows.  In Section~2, we
describe our numerical methods, radiative cooling model, and initial
conditions.  In Section~3 we present our simulation results, focusing
on the jet evolution and induced star formation by the jet bow shock.
These results are discussed further in Section~4, and conclusions are
given in Section~5.

\section{Numerical Simulations}

\subsection{Numerical Methods}

We applied the WENO (weighted essentially non-oscillatory) method (Shu
1999)---a modern high-order upwind method---to the equations of gas
dynamics with atomic and molecular radiative cooling to simulate the
AGN jet and its impact with nearby molecular clouds.  The equations of
gas dynamics with radiative cooling take the form of hyperbolic
conservation laws for mass, momentum, and energy:
\begin{equation}
	\frac{\partial \rho}{\partial t} + 
	\frac{\partial}{\partial x_i} (\rho u_i) = 0
\label{n}
\end{equation}
\begin{equation}
	\frac{\partial}{\partial t} (\rho u_j) + \frac{\partial}{\partial x_i}
	(\rho u_i u_j) + \frac{\partial P}{\partial x_j} = 0
\label{p}
\end{equation}
\begin{equation}
	\frac{\partial E}{\partial t} + \frac{\partial}{\partial x_i}
	\left[u_i (E + P)\right] = - C(n, T) ,
\label{E}
\end{equation}
where $\rho = m n$ is the density of the gas, $m$ is the average mass
of the gas atoms or molecules, $n$ is the number density, $u_i$ is the
velocity, $\rho u_i$ is the momentum density, $P = n k_B T$ is the
pressure, $T$ is the temperature, and $E$ is the energy density of the
gas.  Indices $i,~ j$ equal 1, 2, 3, and repeated indices are summed
over.  The pressure is related to the other state variables by the
equation of state:
\begin{equation}
	P = (\gamma - 1)\left(E - \frac{1}{2} \rho u^2\right) ,
\label{EOS}
\end{equation}
where $\gamma$ is the polytropic gas constant.

We use a ``one fluid'' approximation in the simulations, and assume
that the gas is predominantly H above 8000~K, with the standard
admixture of the other nine most abundant elements He, C, N, O, Ne,
Mg, Si, S, and Fe in the interstellar medium; while below 8000~K, we
assume the gas is predominantly H$_2$, with $n(H)/n(H_2) \approx
0.01$.  We make the further approximation that $\gamma$ is fixed at
$\frac{5}{3}$ (the value for a monatomic gas) for simplicity.

Radiative cooling of the gas is incorporated through the right-hand
side $-C(n, T)$ of the energy conservation equation~(\ref{E}), where
\begin{equation}
        C(n, T) = \left\{ \begin{array}{ll}
        n^2 \Lambda(T) & T \ge 8000~{\rm K,~ for~atomic~cooling~only}\\
        n W(n, T) & T < 8000~{\rm K,~ for~H}_2~{\rm cooling~only}\\
	\end{array} \right.
\label{cooling}
\end{equation}
with the model for $\Lambda(T)$ taken from Figure~8 of Schmutzler \&
Tscharnuter (1993) for atomic cooling, encompassing the relevant
emission lines of the ten most abundant elements in the interstellar
medium, as well as relevant continuum processes; and the model for
$W(n, T)$ from Figure~4 of Le Bourlot et al.\ (1999) for H$_2$
molecular cooling.  Both atomic and molecular cooling are actually
operative between 8000~K $\le T \le$ 10,000~K, but atomic cooling is
dominant in this range.

We use a positivity preserving (Hu et al.\ 2013) version of the
third-order WENO method (Shu 1999, p.~439) for the simulations (Ha et
al.\ 2005).  ENO and WENO schemes are high-order finite difference
schemes designed for nonlinear hyperbolic conservation laws with
piecewise smooth solutions containing sharp discontinuities like shock
waves and contacts.  The WENO schemes use a convex combination of all
candidate upwind and central stencils, rather than just one as in the
original ENO method.

Positivity preserving methods ensure that the gas density and pressure
always remain positive (most numerical methods for gas dynamics can
produce negative densities and pressures) by limiting the numerical
flux.  Without the incorporation of the positivity preserving flux
limiter, the simulations broke down as soon as the jet bow shock
impacted the dense molecular clouds.  

Our WENO code is parallelized as a hybrid MPI/OpenMP program, using
MPI to distribute the computational grid over multiple compute nodes
and localized OpenMP threading to use all cores on a given node.

\subsection{Initial Conditions}

Parallelized simulations were performed on a $420 \Delta x \times 420
\Delta y \times 420 \Delta z$ grid spanning $6 \times 10^{16}$ km =
1.9 kpc on a side, where the grid size $\Delta = \Delta x = \Delta y =
\Delta z$ = 4.6 pc.  The jet was emitted through a disk-shaped inflow
region in the $x y$ plane centered on the $z$-axis with a diameter of
$2 \times 10^{15}$ km, and propagated along the $z$-axis with an
initial velocity of $4 \times 10^{4}$ km s$^{-1}$.  The simulation
parameters at $t = 0$ for the jet, ambient gas, and molecular clouds
are given in Table~\ref{table:params}.  The ambient gas and molecular
clouds were initially pressure-matched with the jet.

\begin{table}[htbp]
\caption{Initial parameters for the jet, ambient gas, and clouds.}
\label{table:params}
\center{
\begin{tabular}{l l l} \hline \hline 
Jet & Ambient Gas & Clouds \\ \hline 
$n_j$ = 0.01 H atoms/cm$^3$ & $n_a = [0.02, 0.2]$ H atoms/cm$^3$ & 
  $n_c = [2, 75]$ H$_2$ molecules/cm$^3$ \\ 
$u_j = 4 \times 10^4$ km s$^{-1}$ & $u_a$ = 0 & $u_c$ = 0 \\ 
$T_j = 10^5$ K & $T_a = [5 \times 10^3, 5 \times 10^4]$ K & 
  $T_c = [7, 250]$ K \\ 
\hline
\end{tabular}
}
\end{table}

AGN jet densities are typically 0.001--0.02 H/cm$^3$ with ambient
interstellar medium densities typically $\sim$ 1 H/cm$^3$ within a
galaxy (Antonuccio et al.\ 2008, Wagner et al.\ 2012).  We take the
average ambient density $\bar{n}_a$ here to be 0.1 H/cm$^3$ as typical
of the less dense {\em intergalactic}\/ medium near the the inner
filament of Cen~A, and take the jet to ambient density ratio
$n_j/\bar{n}_a = 0.1$.  The jet power ${\cal P}_j = 3 \times 10^{43}$
erg s$^{-1}$ is representative of a low power AGN jet like Cen~A: AGN
jets typically have powers between $10^{43}$ and $10^{46}$ erg
s$^{-1}$ (Antonuccio et al.\ 2008, Tortora et al.\ 2009, Wagner et
al.\ 2012).  The jet velocity of $4 \times 10^4$ km s$^{-1}$ = $0.13
c$ gives the correct magnitude of the power as well as the observed
bow shock velocity of the Cen~A jet (Crockett et al.\ 2012).  The
ambient temperature $T_a = 10^4$~K agrees with observations (Crockett
et al.\ 2012), and the jet temperature $T_j = 10^5$~K is chosen to
give pressure-matching between jet and ambient.

\begin{figure}[htbp]
\begin{center}
\includegraphics[scale=0.33]{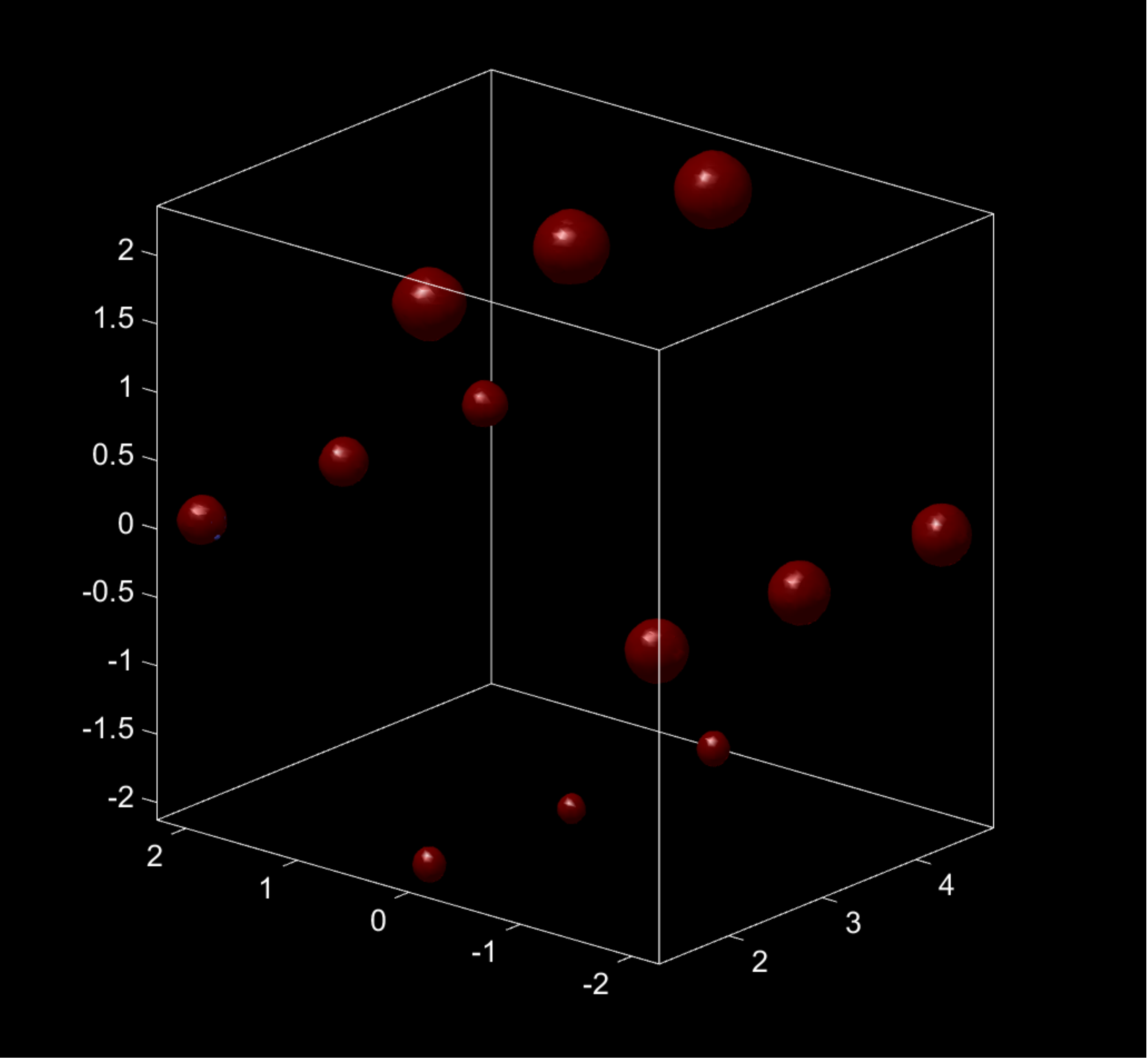}
\caption{Twelve molecular clouds with various diameters (130--325 pc)
  and density ranges are arranged in lines of three centered around
  the jet axis.  Lengths along the boundaries are in units of
  $10^{16}$ km = 324 pc.}
\label{fig:12clouds}
\end{center}
\end{figure}

In order to analyze the consequences of different cloud sizes and
masses, we simulated twelve clouds (arranged in lines of three
centered around the jet axis as shown in Figure~\ref{fig:12clouds})
rather than just the three observed in the inner filament of Cen~A as
imaged by Crockett et al.\ (2012).

\begin{figure}[htbp]
\begin{center}
\includegraphics[scale=0.6]{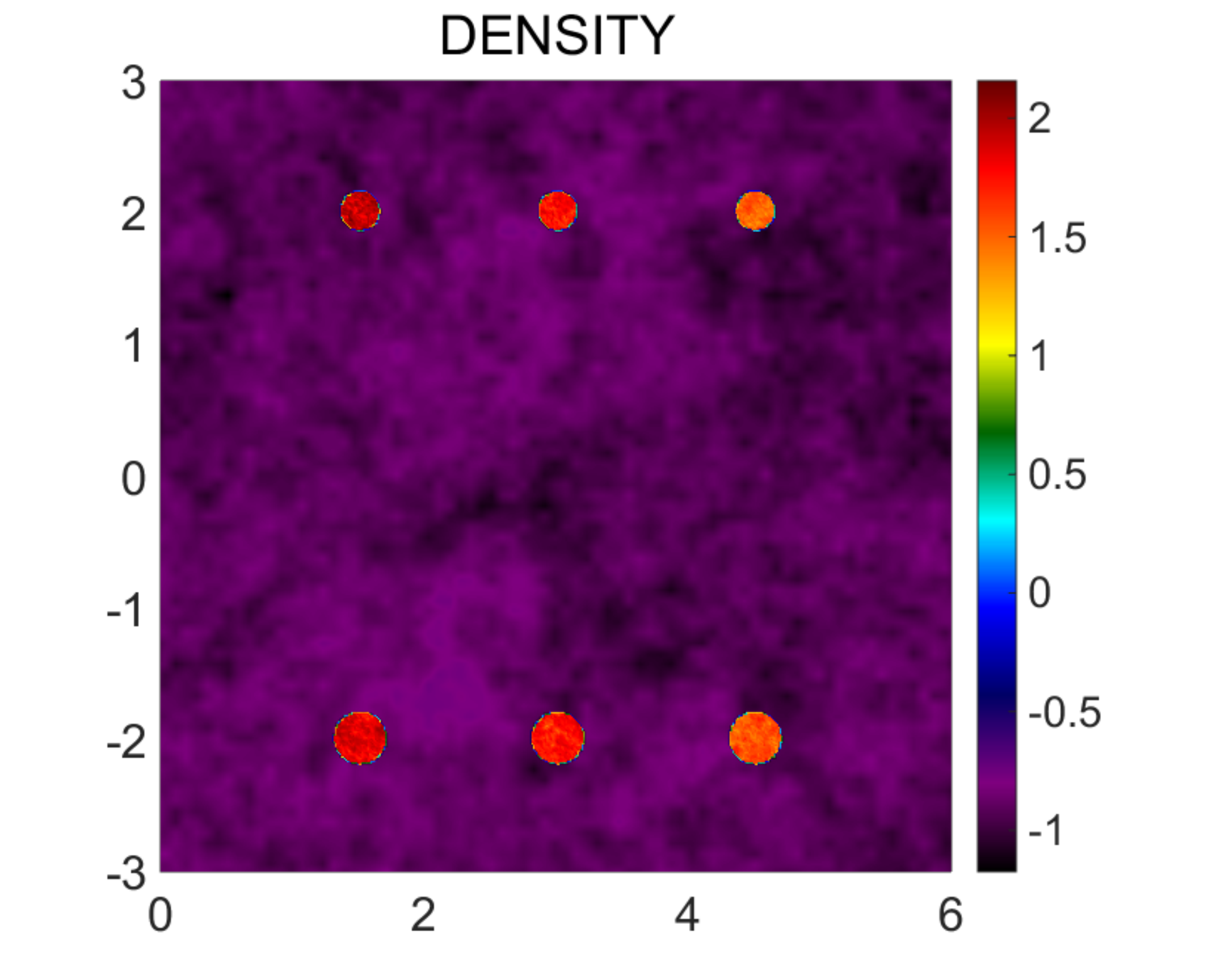}
\caption{Initial values of logarithm of density $\log_{10}(n)$ in
  the $x z$ plane showing the Kolmogorov spectrum of density
  perturbations, with $n$ in H atoms/cm$^3$.  Lengths along the
  boundaries in are in units of $10^{16}$ km = 324 pc.}
\label{fig:rho-init}
\end{center}
\end{figure}

The densities of the ambient gas and the molecular clouds were
initialized (see Figure~\ref{fig:rho-init}) using random perturbations
satisfying the Kolmogorov energy spectrum of a turbulent gas (e.g.,
Elmegreen 2002), within the $[\rho_{\rm min}, \rho_{\rm max}]$ ranges for the
ambient gas and clouds given in Table~\ref{table:params}.  The
physical $(x,y,z)$ grid was filled with random values from a normalized
Gaussian distribution $g$, which was then transformed into Fourier wave
space $(k_x,k_y,k_z)$ to obtain $\hat{g}$.  The transformed Gaussian
was scaled by $\sqrt{k^{-5/3}}$ where $k = (k_x^2+k_y^2+k_z^2)^{1/2}$
to obtain $\hat{f} = \sqrt{k^{-5/3}}\hat{g}$.  The unscaled density
$\tilde{\rho}$ was then calculated as the exponential of the inverse
Fourier transform of $\hat{f}$, i.e., $\tilde{\rho} = \exp(f)$, which
has a log-normal distribution with an energy spectrum satisfying
$|E(k)|^2 \sim k^{-5/3}$.  Finally, the scaled density $\rho$ was
defined by linearly scaling $\tilde{\rho}$ to be in the range
$[\rho_{\rm min},\rho_{\rm max}]$.

\section{Simulation Results}

\subsection{Jet Evolution}

In the simulation figures, the jet is surrounded by a strong bow shock
plus bow-shocked cocoon.  The jet is propagating initially at Mach
3400 with respect to the average sound speed in the ambient gas, and
Mach 1100 with respect to the sound speed in the jet gas.  However,
the jet tip actually propagates at an average velocity of
approximately $1.2 \times 10^4$ km s$^{-1}$, at Mach 1000 with respect
to the average sound speed in the ambient gas, since the jet is slowed
down as it impacts the heterogeneous ambient environment.

The jet is emitted by the AGN at a velocity of approximately $0.3 c$
(Bridle \& Perley 1984), but the late time propagation velocity of the
jet through the lumpy intergalactic medium is impeded down to roughly
$10^4$ km s$^{-1}$.  The transverse propagation velocity of the bow
shock is approximately $10^3$ km s$^{-1}$ (Mach 85 with respect to the
ambient gas), which is comparable to the extrapolated bow shock
velocity of 600--900 km s$^{-1}$ in Crockett et al.\ (2012).

In the plane of the sky, the inner filament of Cen~A lies 8.5 kpc from
the nucleus of NGC 5128, and 2 kpc away from the AGN jet (Morganti et
al. 1999), extending 2200 pc parallel to the jet, with a width of
approximately 50--100 pc.  The initial 500 pc of the inner filament,
beginning in the southwest closest to the galaxy nucleus, comprises
three major groups of extended emission labeled A, B, C in Figure~5 of
Crockett et al.\ (2012).  These three clouds were modeled in the
numerical simulations, with the following caveat: we placed the clouds
at a distance of 650 pc from the jet in order to simulate the impact
of the bow shock on the molecular clouds in a reasonable amount of CPU
time.  The simulations presented here took 100,000 CPU hours on the
SDSC Comet XSEDE computer cluster, using 1200 cores (50 nodes $\times$
24 cores/node).  Moving the clouds further out to 2 kpc away from the
jet and increasing the bow shock impact time with the clouds from 0.7
Myr to approximately 3 Myr would require roughly $3^4 \cdot 5$ times
as long with the same spatial resolution.  Note that we need at least
the current spatial resolution to resolve nascent star clusters, which
encompass here $4^3$ grid cells.  We believe that the main physical
effects are captured in our current simulations, while keeping the CPU
time feasible.

The molecular clouds in the inner filament were initialized to be in
agreement with the estimate of Crockett et al.\ (2012) that the
pre-ablation mass of Regions~1 and~2 in cloud~A is approximately $6
\times 10^5 \, M_\odot$.  We studied a range of initial cloud masses
from $10^5$ to $2 \times 10^6 \, M_\odot$, with diameters 130--325 pc.

\subsection{Star Formation}

We use the three criteria of Cen \& Ostriker (1992) to test for star
formation, applied to averaged values in a spherical region of radius
$R$:
\begin{equation}
        \div {\bf u} < 0 ,
\label{i}
\end{equation}
\begin{equation}
        t_{\rm cool} = \frac{E}{|dE/dt|} = \frac{E}{C} \ll t_{\rm grav} = 
        \pi^\half G^{-\half} \rho^{-\half} ,
\label{ii}
\end{equation}
\begin{equation}
        M_{\rm region} = \frac{4 \pi}{3} \rho R^3 > M_{\rm Je} \equiv
        \frac{4 \pi}{3} \rho \lambda_{\rm Je}^3 \approx 
        G^{-\3half} \rho^{-\half} c_s^3 ~~~{\rm (Jeans~unstable)} ,
\label{iii}
\end{equation}
where $t_{\rm cool}$ is the characteristic radiative cooling time,
$t_{\rm grav}$ is the characteristic gravitational collapse time,
$M_{\rm Je}$ is the Jeans mass, $\lambda_{\rm Je} \equiv c_s t_{\rm
  grav}$ is the Jeans length, and $c_s = \sqrt{\gamma P/\rho} =
\sqrt{\gamma k_B T/m}$~ is the sound speed.  We define the Jeans ratio
as ${\rm Je} \equiv M_{\rm region}/M_{\rm Je}$.  In
equation~(\ref{iii}), ${\rm Je} > 1$ means the region is Jeans
unstable, i.e., the gravitational free-fall time is shorter than the
time needed for a pressure wave to traverse the collapsing region and
bounce back, resisting collapse.

Thus the region will collapse gravitationally if ${\rm Je} > 1$ and
$t_{\rm cool} \ll t_{\rm grav}$, plus $\div {\bf u} < 0$, which is
typically true in bow-shocked regions.  Note that equations~(\ref{ii})
and~(\ref{iii}) are equivalent to
\begin{equation}
        \frac{t_{\rm cool}}{t_{\rm grav}} \approx 
        \frac{E G^\half \rho^\half}{C} \ll 1 ,
\label{iia}
\end{equation}
and
\begin{equation}
        {\rm Je} = \left(\frac{R}{\lambda_{\rm Je}}\right)^3 \approx
        R^3 G^\3half \rho^\3half c_s^{-3} \approx 
        R^3 G^\3half m^\3half \left(\frac{\rho}{k_B T}\right)^\3half > 1 ,
\label{iiia}
\end{equation}
respectively.  Thus star formation is facilitated by a simultaneous
combination of high densities and low temperatures from
equation~(\ref{iiia}), {\em plus}\/ strong radiative cooling---which
requires $T$ not to be too low---from equation~(\ref{iia}).

To identify star cluster forming regions, we first divided the
computational grid into cubic cells of length $D$ on a side, where $D$
= 20 pc = $4 \Delta$ is the typical diameter of a nascent cluster.
Next we averaged all state variables and $\div {\bf u}$ within each
cell, and for each averaged cell, we computed the mass within the cell
$M_{\rm cell} \equiv M_{\rm region}$, the Jeans mass $M_{\rm Je}$, the
cooling time $t_{\rm cool}$, and the gravitational free-fall time
$t_{\rm grav}$.  If the three conditions above for star formation held
for an averaged cell (we used $t_{\rm cool} \le 0.2 \, t_{\rm grav}$),
we marked that cell as a star cluster.

For the molecular clouds considered here, $D$ = 20 pc is the
physically relevant length scale in that for $D$ = 10 pc many adjacent
star clusters form which should be merged into a single star cluster,
and for $D$ = 30 pc some star forming regions are suppressed by the
averaging with neighboring regions that do not form stars.  The
computed post-ablation averaged quantities in the star clusters are
given in Table~\ref{table:clusters20}, with 71 clusters identified in
the twelve clouds when $D$ = 20 pc.

\begin{table}[htbp]
\caption{Computed ranges for averaged values in the star cluster cells
  at 780,000 yr, with 71 clusters identified in the twelve clouds when
  the cluster diameter $D$ = 20 pc.}  \center{
\label{table:clusters20}
\begin{tabular}{l l l l} \hline \hline
Quantity & Min Value & Max Value & Units \\ \hline
$n$  & $32$ & $60$ & H$_2$ molecules/cm$^3$ \\
$T$     & $15$ & $270$ & K \\
$c_s$   & $0.45$ & $1.9$ & km s$^{-1}$ \\
$t_{\rm cool}$   & $0.03$ & $1.6$ & Myr \\
$t_{\rm grav}$   & $8.7$ & $12$ & Myr \\
$t_{\rm cool}/t_{\rm grav}$ & $0.003$ & $0.15$ & \\
$\lambda_{\rm Je}$ & $4.3$ & $18$ & pc \\
$M_{\rm cluster}$ & $10^4$ & $1.9 \times 10^4$ & $M_\odot$ \\
$M_{\rm Je}$ & $200$ & $1.6 \times 10^4$ & $M_\odot$ \\
${\rm Je}$ & $1.1$ & $77$ & \\
\hline
\end{tabular}
}
\end{table}

\section{Discussion}

\begin{figure}[htbp]
\begin{center}
\includegraphics[scale=0.95]{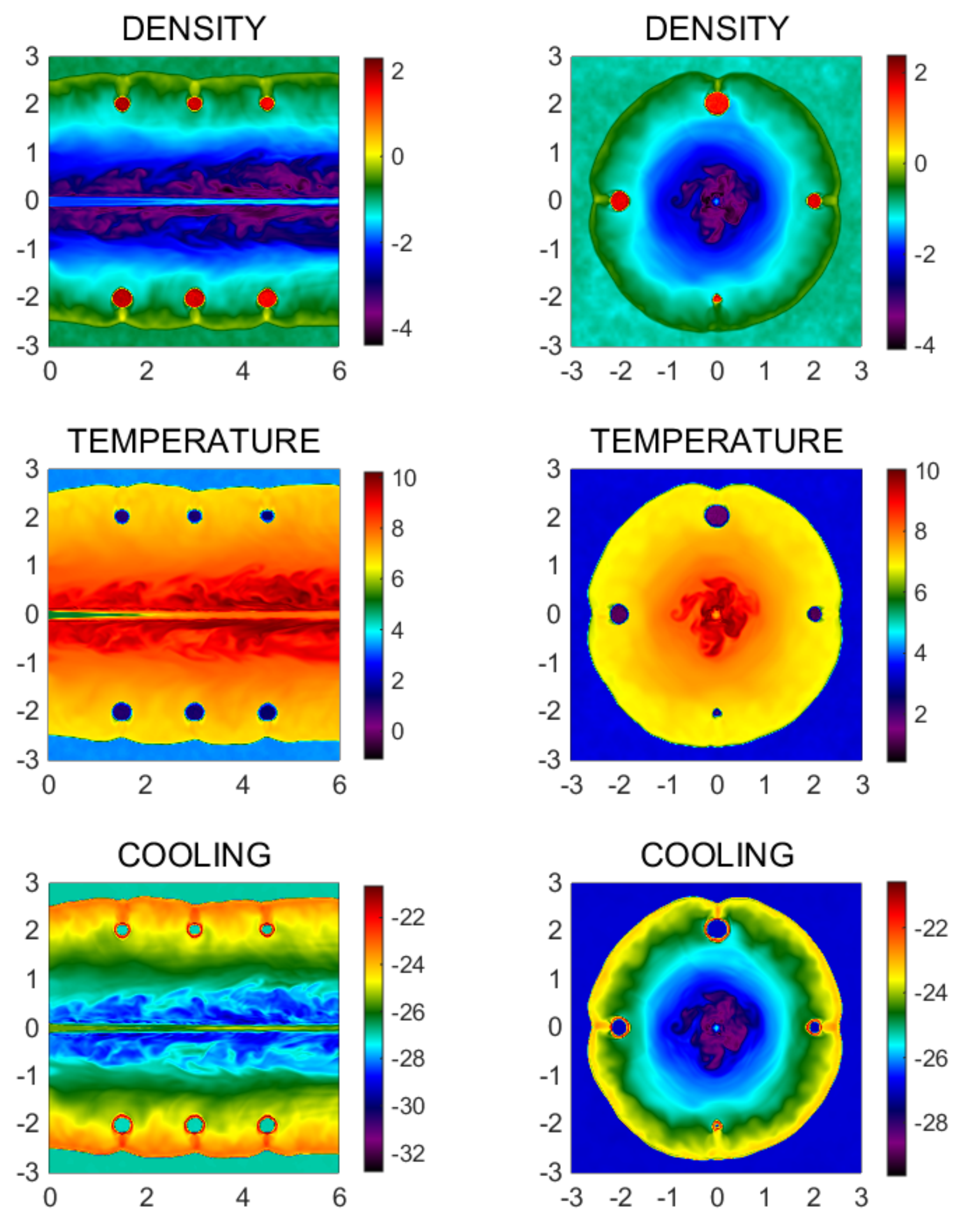}
\caption{Logarithm of density $\log_{10}(n)$ with $n$ in H
  atoms/cm$^3$, logarithm of temperature $\log_{10}(T)$ with $T$ in~K,
  and logarithm of total radiative cooling $\log_{10}(C)$ with $C$ in
  erg cm$^{-3}$ s$^{-1}$ at $t = 780,000$ yr.  Left panel: in the $x
  z$ plane; right panel: in the central $z$-slice.  Lengths along the
  boundaries are in units of $10^{16}$ km = 324 pc.}
\label{fig:combo-all}
\end{center}
\end{figure}

Cross sections of the 3D numerical simulations at 780,000 yr---just
after the bow shock has traversed the molecular clouds---are shown in
Figure~\ref{fig:combo-all}.  Turbulent Kelvin-Helmholtz mixing of the
jet and ambient gas is evident in a layer of roughly $2 \times
10^{16}$ km around the jet.  This layer has the lowest densities $n
\approx 10^{-3.5}$ H atoms/cm$^3$ and highest temperatures $T \approx
10^8$--$10^{10}$~K; it also has a strong backflow counter to the
direction of the jet flow.  Post-shock ablation of the clouds is also
clearly visible in Figure~\ref{fig:combo-all}.  Even though the clouds
are at the lowest temperatures $T \approx 15$--270~K, the strongest
radiative cooling (which is essential for star formation) takes place
around the surfaces of the clouds.  While the cloud cores are not
cooling as intensely, they are at lower temperatures and higher
densities than the surfaces, so star clusters form in the cloud cores
as well.  It is the combination of high densities, low temperatures,
{\em and}\/ strong radiative cooling that enables star formation.

The age of the inner lobes of Cen~A is 5.6--5.8 Myr and the inner
filament stars are roughly 1--3 Myr old (Crockett et al.\ 2012).
Radio and X-ray synchrotron source lifetimes are in general on the
order of 10~Myr, and therefore comfortably longer than the star
formation times.  The AGN jet thus exists long enough to trigger star
formation in nearby molecular clouds which are large enough and dense
enough; in our simulations, star formation is triggered by 780,000 yr.

Table~\ref{table:Nclusters} shows the initial cloud parameters and the
number of star clusters formed at 780,000 yr in each cloud.  What is
evident from the data is that more than one star cluster forms for $D$
= 20 pc only if $n_{\rm max} \ge 50$ H$_2$ molecules/cm$^3$.  For
$D$ = 20 pc, five out of twelve clouds form a significant number of
star clusters.  However, recall that in the interstellar medium on
average only about 10\% of regions satisfying the star formation
criteria actually collapse to form star clusters.  What is important
for our purposes is that there are a significant number of clouds that
do not form star clusters, while the denser clouds do.

\begin{table}[htbp]
\caption{Initial cloud parameters and the number $N_{\rm 20 pc}$ of
  star clusters for $D$ = 20 pc formed at 780,000 yr in each cloud,
  with $\bar{n}$ = 1 H$_2$ molecule/cm$^3$.  For all clouds,
  $n_{\rm min} = 2$ H$_2$ molecules/cm$^3$.  Rows are ranked by
  $N_{\rm 20 pc}$.}  \center{
\label{table:Nclusters}
\begin{tabular}{c c c c c} \hline \hline
Cloud & $n_{\rm max}/\bar{n}$ & $D_{\rm cloud}/$pc & 
  $M_{\rm cloud}$/($10^5$ $M_\odot)$ & $N_{\rm 20 pc}$ \\ 
\hline
1  & 60 & 260 & 17 & 29 \\
2  & 75 & 195 & 9  & 18 \\
3  & 50 & 195 & 6  & 10 \\
4  & 50 & 260 & 15 & 7  \\
5  & 75 & 130 & 3  & 4  \\
6  & 50 & 130 & 2  & 2  \\
7  & 40 & 325 & 23 & 1  \\
8  & 30 & 325 & 18 & 0  \\
9  & 30 & 260 & 9  & 0  \\
10 & 25 & 195 & 3  & 0  \\
11 & 25 & 130 & 1  & 0  \\
12 & 20 & 325 & 12 & 0  \\
\hline 
\end{tabular}
}
\end{table}

\begin{figure}[htbp]
\begin{center}
\includegraphics[scale=0.4]{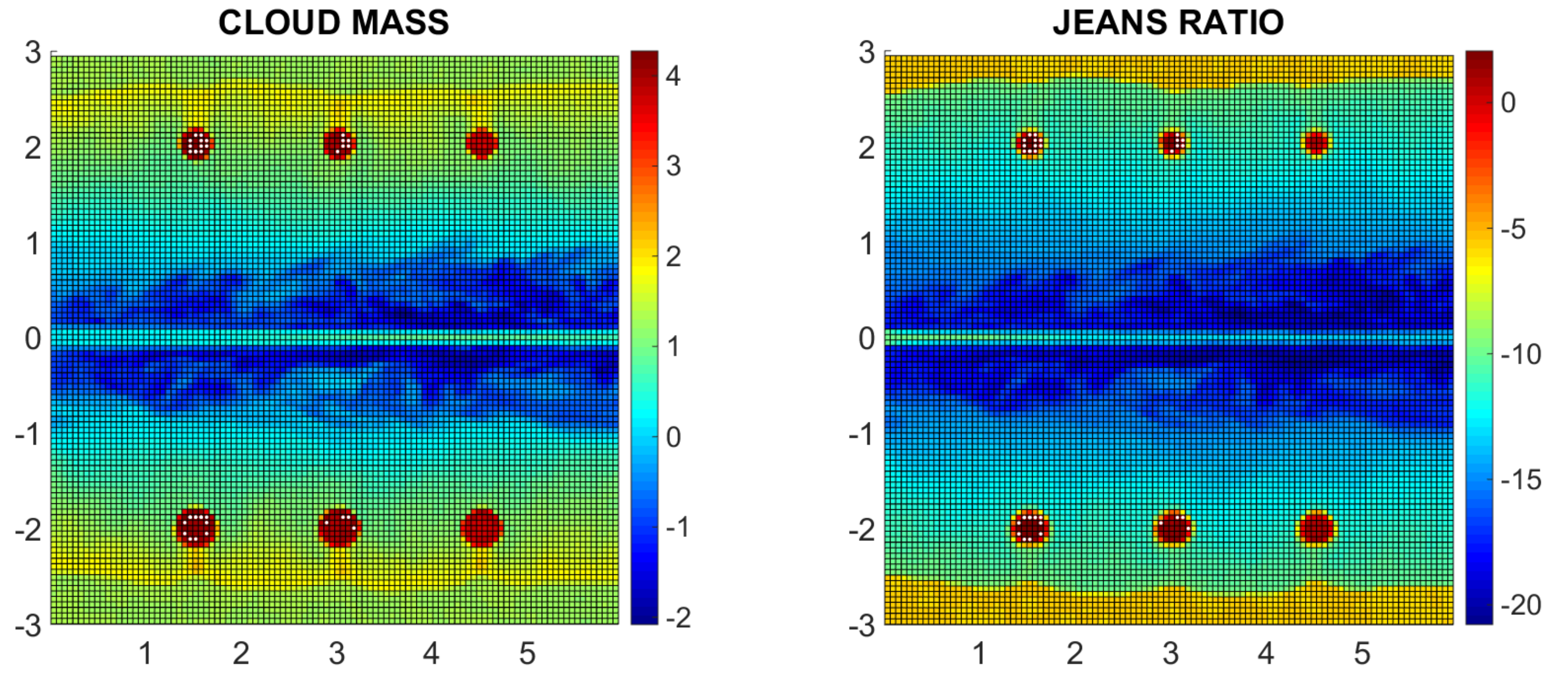}
\caption{Left panel: $\log_{10}(M_{\rm cell}/M_\odot)$ in each cell,
  with $D$ = 20 pc. Right panel: $\log_{10}({\rm Je})$ in each cell.
  Regions satisfying the star formation criteria are marked with white
  dots.  Lengths along the boundaries are in units of $10^{16}$ km =
  324 pc.}
\label{fig:cells}
\end{center}
\end{figure}

\begin{figure}[htbp]
\begin{center}
\includegraphics[scale=0.4]{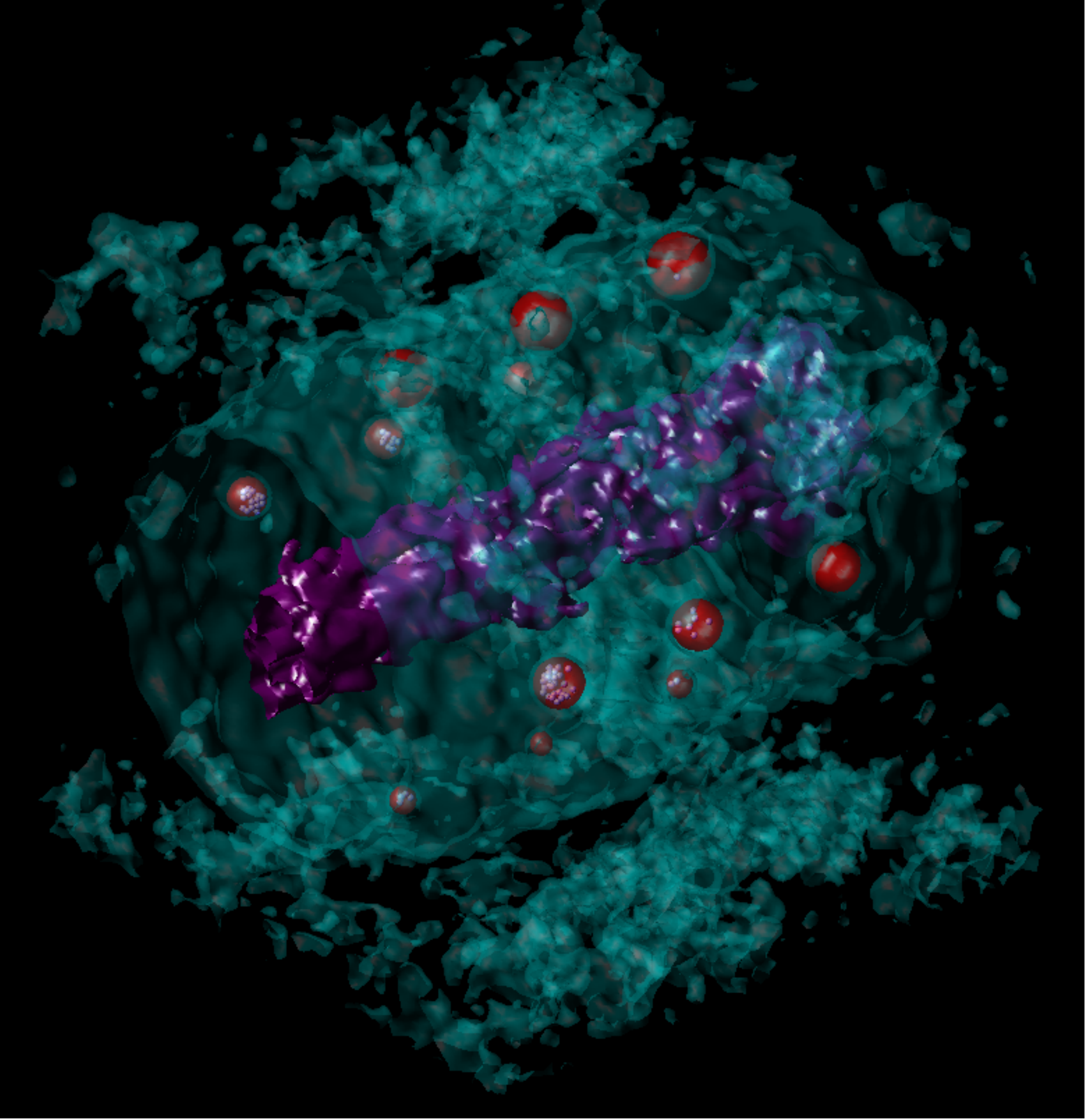}
\caption{Density isosurfaces at $t = 780,000$ yr.  Purple: low density
  ($n = 10^{-3}$ H atoms/cm$^3$) jet cocoon; cyan: medium density
  ($n = 0.1$ H atoms/cm$^3$) ambient and bow-shocked regions; red:
  high density ($n = 5$ H$_2$ molecules/cm$^3$) molecular clouds.
  Star clusters for $D$ = 20 pc are shown as white spheres.}
\label{fig:3D-dens}
\end{center}
\end{figure}

\begin{figure}[htbp]
\begin{center}
\includegraphics[scale=0.5]{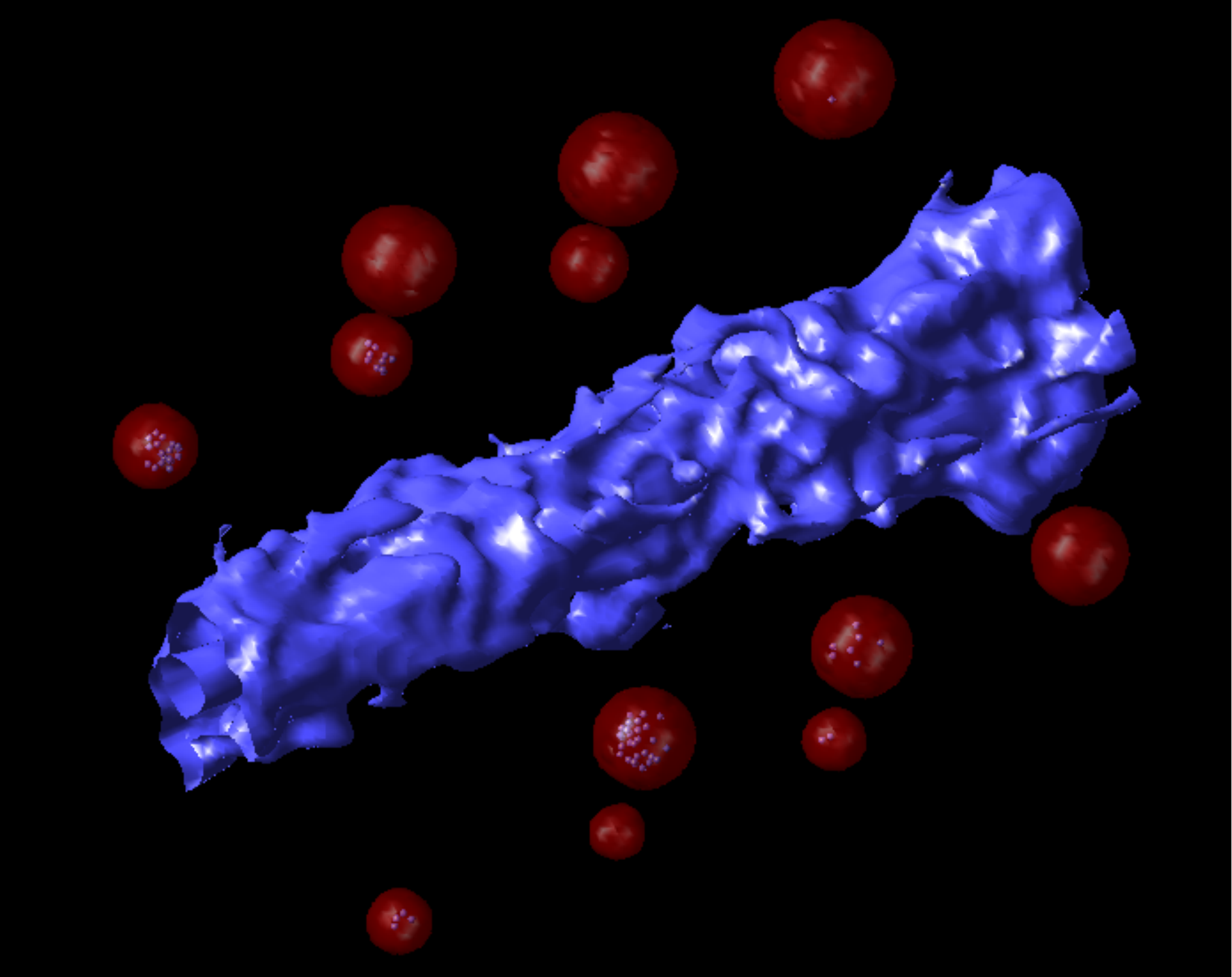}
\caption{Density isosurfaces plus star clusters shown as white
  spheres.  71 clusters of mass 1--$1.9 \times 10^4 \, M_\odot$ were
  found with $D$ = 20 pc.}
\label{fig:stars20-10}
\end{center}
\end{figure}

Figure~\ref{fig:cells} shows the $xz$ plane slice of the cells used to
identify star cluster regions for $D$ = 20 pc.  The locations of the
nascent star clusters are indicated in Figures~\ref{fig:3D-dens}
and~\ref{fig:stars20-10}, with 71 clusters of mass $M_{\rm cluster} =
1$--$1.9 \times 10^4 \, M_\odot$ formed with $D$ = 20 pc.  The
simulated clouds with approximately 20 clusters are a good match to
cloud~A in the {\em HST}\/ images, while clouds~B and~C show no
clusters, matching several of our simulated clouds.

\begin{figure}[htbp]
\begin{center}
\includegraphics[scale=0.6]{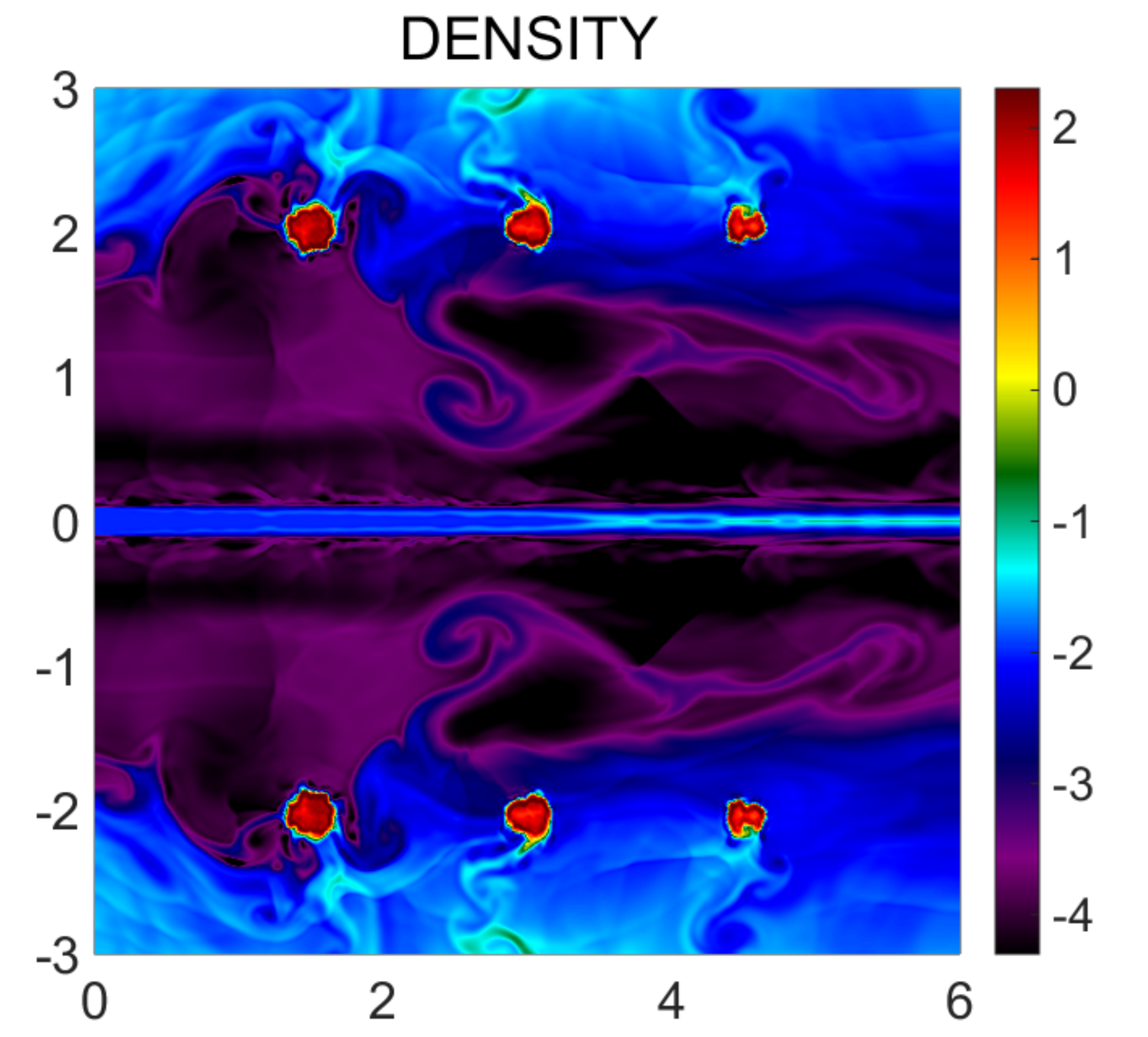}
\caption{Logarithm of density $\log_{10}(n)$ in the $r z$ plane at
  $t$ = 3 Myr for a cylindrically symmetric simulation, with $n$ in
  H atoms/cm$^3$.  Lengths along the boundaries are in units of
  $10^{16}$ km = 324 pc.}
\label{fig:cyl-sym}
\end{center}
\end{figure}

The star clusters in Figure~\ref{fig:stars20-10} have already
persisted to roughly $t_{\rm grav}/10$.  Persistence of the star
clusters to late times on the order of $t_{\rm grav}/3 \approx$ 3 Myr
was verified by cylindrically symmetric simulations with exactly the
same initial parameters as in the 3D simulations and with a final time
of 3 Myr.  The greater density contrast evident in
Figure~\ref{fig:cyl-sym} is due to the much further development of
Kelvin-Helmholtz vortex rollup inside the jet cocoon.  At 1 Myr there
are 27 star clusters in the three spherical clouds (considered as part
of the cylindrically symmetric cloud ``ring'' near the $r z$ cross
section), while at 3 Myr there are 30 clusters, so the number of star
clusters is actually increasing slightly between 1 Myr and 3 Myr.

\begin{figure}[htbp]
\begin{center}
\includegraphics[scale=0.5]{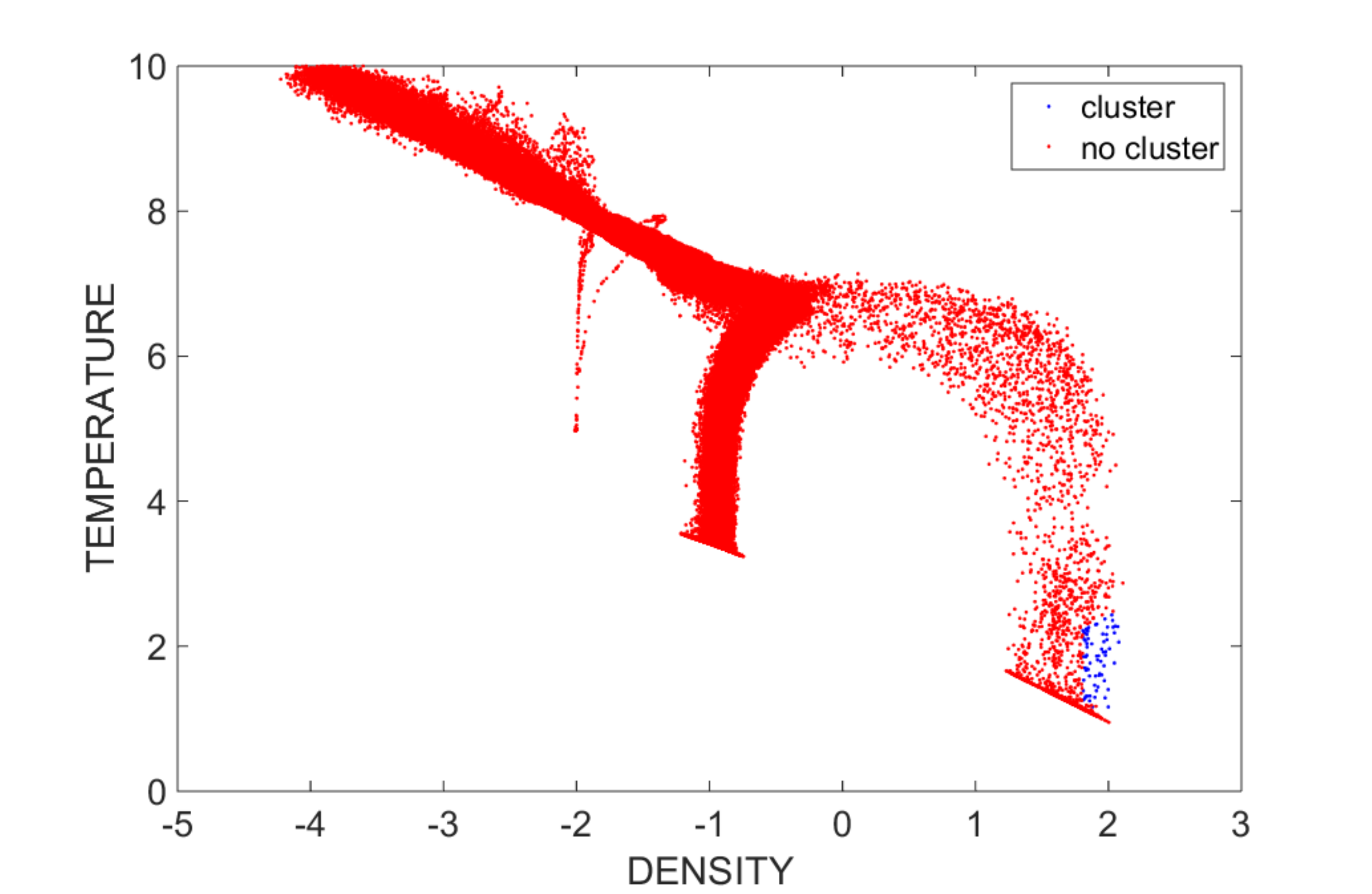}
\caption{$\log_{10}(T)$ vs.\ $\log_{10}(n)$ for averaged cell
  values with $D$ = 20 pc, with $T$ in~K and $n$ in H atoms/cm$^3$.
  The star cluster regions require high density and low temperature,
  and are shown by blue dots.}
\label{fig:scatter}
\end{center}
\end{figure}

In Figure~\ref{fig:scatter}, a plot of $\log_{10}(T)$
vs.\ $\log_{10}(n)$ is shown for averaged cell values from the 3D
simulations, demonstrating that star cluster regions require high
density and low temperature.  However if the temperature is too low,
clusters do not form because the radiative cooling is too weak and
$t_{\rm cool} \approx t_{\rm grav}$ (lower right-hand corner of the
scatter plot).

Several regions can be clearly seen in the figure.  The inner cocoon
points can be seen in the ``handle'' shape in the upper left-hand part
of the plot, with high temperatures and low densities.  The points
near the jet inflow can be seen as deviations from the handle near
$n = n_j = 10^{-2}$ H atoms/cm$^3$.  Points in the outer
parts of the cocoon and in the unshocked ambient region lie in the
dense vertical strip near the average ambient density $n \approx
10^{-1}$ H atoms/cm$^3$.  The more sparsely scattered points in the
rightmost region are in the molecular clouds, with the star-forming
regions in the lower right-hand corner, with low temperatures and high
densities.

Since $P = n k_B T$, the log-log plot of a pressure-matched gas would
be a straight line with constant slope and a $y$-intercept of $\log
P$.  Two main trend lines are visible in the figure, showing two
disparate pressure regimes.  The lower pressure line corresponds to
the original pressure of the entire region $P_j = P_a = P_c \approx
0.1 $ eV cm$^{-3}$, which includes all of the ambient points and most
of the points in the molecular clouds.  The higher pressure line
corresponds to a pressure four orders of magnitude greater than $P_j$,
and contains mainly the bow-shocked cocoon points.  Although it is not
apparent from the plot because many data points are plotted over each
other, the data points lying between the two pressure regimes make up
only a small fraction of the total number of points and correspond to
points near the boundaries of the cocoon or clouds.

\section{Conclusion}

The 3D numerical simulations demonstrate that the transverse bow shock
of the Cen~A jet induces star formation as it impacts surrounding
dense molecular clouds.  The passage of the bow shock causes the
molecular clouds to undergo strong radiative cooling, thereby
triggering the star formation observed in the inner filament of
Cen~A\@.  We find that star clusters form inside a bow-shocked
molecular cloud when the maximum initial density of the cloud is $\ge
40$ H$_2$ molecules/cm$^3$.  In a typical molecular cloud of mass
$10^6 \, M_\odot$ and diameter 200 pc, approximately 20 star clusters
of mass $10^4 \, M_\odot$ are formed, matching the {\em HST}\/ images
of cloud~A in Crockett et al.\ (2012).

Cross sections of the 3D numerical simulations at 780,000 yr---just
after the bow shock has traversed the molecular clouds---display
turbulent Kelvin-Helmholtz mixing of the jet and ambient gas in a
layer of roughly $2 \times 10^{16}$ km around the jet.  This layer has
the lowest densities $n \approx 10^{-3.5}$ H atoms/cm$^3$ and
highest temperatures $T \approx 10^8$--$10^{10}$~K; it also has a
strong backflow counter to the direction of the jet flow.  Post-shock
ablation of the clouds is also clearly visible.  Even though the
clouds are at the lowest temperatures $T \approx 15$--270~K, the
strongest radiative cooling---which is essential for star
formation---takes place around the surfaces of the clouds.  While the
cloud cores are not cooling as intensely, they are at lower
temperatures and higher densities than the surfaces, so star clusters
form in the cloud cores as well.  It is the combination of high
densities, low temperatures, {\em and}\/ strong radiative cooling that
enables star formation.  However if the temperature is too low,
clusters do not form because the radiative cooling is too weak and
$t_{\rm cool} \approx t_{\rm grav}$.

In future work, we plan to include gravity in the 3D gas dynamical
solver, and to follow a collapsing cloud with approximately 20 nascent
star clusters at 780,000 yr out to $t_{\rm grav} \approx$ 10 Myr.
Close examination of the composite image of Cen~A in
Figure~\ref{fig:CenAStars} shows a line of blue star-forming regions,
stretching from just below the tip of the X-ray jet all the way out to
the inner filament, along the transverse bow shock of the jet.  These
star-forming regions, which we plan to model in detail in future work,
would make excellent targets for further observations with the {\em
  HST WFC3}.

\acknowledgments

ES was supported by NSF grant AST14-07835 and NASA theory grant
NNX15AK82G\@.  We would like to thank the San Diego Supercomputer
Center (SDSC) at the University of California, San Diego, and the
Extreme Science and Engineering Discovery Environment (XSEDE) for
providing HPC resources via grant TG-AST130021.  RAW acknowledges
support from HST grant GO-11359 from the Space Telescope Science
Institute, which is operated by the Association of Universities for
Research in Astronomy, Inc., under NASA contract NAS 5-26555.  RAW
also acknowledges support from NASA JWST Interdisciplinary Scientist
grants NAG5-12460 and NNX14AN10G from GSFC.

\section*{References}

\begin{itemize}

\setlength\itemsep{0em}

\item[] Antonuccio-Delogu, V., \& Silk, J.\ 2008, MNRAS, 389, 1750

\item[] Bicknell, G.~V., Sutherland, R.~S., van Breugel, W.~J.~M., et
  al.\ 2000, ApJ, 540, 67

\item[] Bridle, A.~H., \& Perley, R.~A. 1984, ARA\&A, 22, 319

\item[] Brinchmann, J., \& Ellis, R.~S.\ 2000, ApJ, 536, L77

\item[] Cavagnolo, K.~W., Donahue, M., Voit, G.~M., \& Sun, M.\ 2009,
  ApJS, 182, 12

\item[] Cen, R., \& Ostriker, J.~P. 1992, ApJ, 399, L113

\item[] Costa, T., Sijacki, D., \& Haehnelt, M.~G.\ 2014, MNRAS, 444,
  2355

\item[] Cowie, L.~L., \& Barger, A.~J.\ 2008, ApJ, 686, 72

\item[] Cowie, L.~L., Songaila, A., Hu, E.~M., \& Cohen, J.~G.\ 1996,
  AJ, 112, 839

\item[] Crockett, R.~M., Shabala, S.~S., Kaviraj, S., et al.\ 2012,
  MNRAS, 421, 1603

\item[] Croft, S., van Breugel, W., de Vries, W., et al.\ 2006, ApJ,
  647, 1040

\item[] Croton, D.~J., Springel, V., White, S.~D.~M., et al.\ 2006,
  MNRAS, 365, 11

\item[] Dey, A., van Breugel, W., Vacca, W.~D., \& Antonucci,
  R.\ 1997, ApJ, 490, 698

\item[] Dugan, Z., Bryan, S., Gaibler, V., Silk, J., \& Haas, M. 2014,
  ApJ, 796, 113

\item[] Dugan, Z., Gaibler, V., Silk, J. 2016, arXiv:1608.01370

\item[] Dunn, R.~J.~H., \& Fabian, A.~C.\ 2006, MNRAS, 373, 959

\item[] Elmegreen, B.~G. 2002, ApJ, 564, 773

\item[] Fabian, A.~C.\ 2012, ARA\&A, 50, 455 

\item[] Fassett, C.~I., \& Graham, J.~A.\ 2000, ApJ, 538, 594

\item[] Feain, I.~J., Ekers, R.~D., Murphy, T., et al.\ 2009, ApJ,
  707, 114

\item[] Gaibler, V. 2014, AN, 335, 531

\item[] Gaibler, V., Khochfar, S., Krause, M., \& Silk, J. 2012,
  MNRAS, 425, 438

\item[] Gardner, C.~L., \& Dwyer, S.~J. 2009, AcMaS, 29B, 1677

\item[] Gaspari, M., Brighenti, F., \& Temi, P.\ 2012a, MNRAS, 424,
  190

\item[] Gaspari, M., Ruszkowski, M., \& Sharma, P.\ 2012b, ApJ, 746,
  94

\item[] Gopal-Krishna, \& Wiita, P.~J.\ 2010, NA, 15, 96

\item[] Granato, G.~L., De Zotti, G., Silva, L., Bressan, A., \&
  Danese, L.\ 2004, ApJ, 600, 580

\item[] Graham, J.~A., \& Fassett, C.~I.\ 2002, ApJ, 575, 712

\item[] Ha, Y., Gardner, C.~L., Gelb, A., \& Shu, C.-W. 2005, JSCom,
  24, 29

\item[] Hu, X.~Y., Adams, N.~A., \& Shu, C.-W. 2013, JCP, 242, 169

\item[] Israel, F.~P.\ 1998, ARA\&A, 8, 237 

\item[] Hopkins, A.~M., \& Beacom, J.~F.\ 2006, ApJ, 651, 142 

\item[] Hopkins, P.~F., Hernquist, L., Cox, T.~J., et al.\ 2006,
  ApJS, 163, 1

\item[] Johansson, P.~H., Naab, T., \& Ostriker, J.~P.\ 2012, ApJ,
  754, 115

\item[] Karachentsev, I. D.  et al.\ A\&A, 2002, 385, 21

\item[] Karim, A., Schinnerer, E., Mart{\'{\i}}nez-Sansigre, A., et
  al.\ 2011, ApJ, 730, 61

\item[] Kraft, R.~P., Forman, W.~R., Hardcastle, M.~J., et al.\ 2009,
  ApJ, 698, 2036

\item[] Le Bourlot, J., Pineau des For\^ets, G., \& Flower,
  D.~R. 1999, MNRAS, 305, 802

\item[] Li, Y., Bryan, G.~L., Ruszkowski, M., et al.\ 2015, ApJ, 811,
  73

\item[] McNamara, B.~R., \& Nulsen, P.~E.~J.\ 2007, ARA\&A, 45, 117

\item[] Minkowski, R.\ 1958, PASP, 70, 143

\item[] Morganti, R., Killeen, N.~E.~B., Ekers, R.~D., Oosterloo,
  T.~A. 1999, MNRAS, 307, 750

\item[] Morganti, R., Robinson, A., Fosbury, R.~A.~E., et al.\ 1991,
  MNRAS, 249, 91

\item[] Noeske, K.~G., Faber, S.~M., Weiner, B.~J., et al.\ 2007,
  ApJ, 660, L47

\item[] Peterson, J.~R., Paerels, F.~B.~S., Kaastra, J.~S., et
  al.\ 2001, A\&A, 365, L104

\item[] Rafferty, D.~A., McNamara, B.~R., Nulsen, P.~E.~J., \& Wise,
  M.~W.\ 2006, ApJ, 652, 216

\item[] Rejkuba, M., Greggio, L., \& Zoccali, M.\ 2004, A\&A, 415, 915

\item[] Rejkuba, M., Minniti, D., Courbin, F., \& Silva, D.~R.\ 2002,
  ApJ, 564, 688

\item[] Rejkuba, M., Minniti, D., Silva, D.~R., \& Bedding,
  T.~R.\ 2001, A\&A, 379, 781

\item[] Richardson, M.~L.~A., Scannapieco, E., Devriendt, J., et
  al.\ 2016, ApJ, 825, 83

\item[] Saxton, C.~J., Sutherland, R.~S., \& Bicknell, G.~V.\ 2001,
  ApJ, 563, 103

\item[] Scannapieco, E., \& Oh, S.~P.\ 2004, ApJ, 608, 62

\item[] Scannapieco, E., Silk, J., \& Bouwens, R.\ 2005, ApJ, 635,
  L1

\item[] Schmutzler, T., \& Tscharnuter, W.~M. 1993, A\&A, 273, 318

\item[] Shu, C.-W. 1999, High-Order Methods for Computational Physics, Vol.~9
(New York: Springer)

\item[] Silk, J. 2013, ApJ, 772, 112

\item[] Somerville, R.~S., Hopkins, P.~F., Cox, T.~J., Robertson,
  B.~E., \& Hernquist, L.\ 2008, MNRAS, 391, 481

\item[] Sutherland, R.~S., Bicknell, G.~V., \& Dopita, M.~A. 1993,
  ApJ, 414, 510

\item[] Tortora, C., Antonuccio-Delogu, V., Kaviraj, S., Silk, J.,
  Romeo, A.~D., \& Becciani, U.\ 2009, MNRAS, 396, 61

\item[] van Breugel, W., Filippenko, A.~V., Heckman, T., \& Miley,
  G.\ 1985, ApJ, 293, 83

\item[] Wagner, A.~Y., Bicknell, G.~V., \& Umemura, M.\ 2012, ApJ, 757, 136

\item[] Wagner, A.~Y., Bicknell, G.~V., Umemura, M., Sutherland,
  R.~S., \& Silk, J. 2016, AN, 337, 167

\item[] Wurster, J., \& Thacker, R.~J.\ 2013, MNRAS, 431, 2513

\item[] Zinn, P.-C., Middelberg, E., Norris, R.~P., \& Dettmar,
  R.-J. 2013, ApJ, 774, 66

\end{itemize}

\end{document}